# Designing the Future of Entrepreneurship Education: Exploring an AI-Empowered Scaffold System for Business Plan Development[1]

ZHU Junhua
*College of Education Sciences*
Hong Kong University of Science and Technology (Guangzhou)
Guangzhou, China
junhuazhu@hkust-gz.edu.cn

LUO Lan*
*Computation Media and Arts*
Hong Kong University of Science and Technology (Guangzhou)
Guangzhou, China
lluo476@connect.hkust-gz.edu.cn

*Abstract*—Entrepreneurship education equips students to transform innovative ideas into actionable entrepreneurship plans, yet traditional approaches often struggle to provide the personalized guidance and practical alignment needed for success. Focusing on the business plan as a key learning tool and evaluation method, this study investigates the design needs for an AI-empowered scaffold system to address these challenges. Based on qualitative insights from educators and students, the findings highlight three critical dimensions for system design: mastery of business plan development, alignment with entrepreneurial learning goals, and integration of adaptive system features. These findings underscore the transformative potential of AI in bridging gaps in entrepreneurship education while emphasizing the enduring value of human mentorship and experiential learning.

## I. Introduction

Entrepreneurship education is a cornerstone of higher education, crucial for fostering innovation, driving economic growth, and equipping students to navigate complex real-world challenges [1]. While its precise definition remains contested [2], it is broadly understood as the process of developing students' mindset, skills, and competencies—not only for launching ventures but also for applying entrepreneurial thinking across diverse sectors [3][4]. Entrepreneurship education should be conceptualized as a scaffolding continuum, progressing from innovation, which centers on the generation of new ideas, to entrepreneurship, which focuses on creating tangible value and profit [5]. In this context, the business plan is often employed as a classic tool for assessing students' entrepreneurial mindset and the skills they develop in the classroom [3]. It embodies the core logic of entrepreneurship education by scaffolding students' understanding of entrepreneurial principles and providing a structured framework for articulating their ideas and strategies.

Despite its importance, both business plan development and broader entrepreneurship education face several common challenges. These include an overemphasis on theoretical content, limited interaction with real-world practices, and a lack of personalized learning experiences [6] [7]. Curricula are often redundant and lack sufficient differentiation, preventing them from adequately addressing the diverse needs of students [8]. Teaching methods tend to prioritize lectures, with insufficient emphasis on practical application and hands-on learning [2]. Furthermore, there is a shortage of qualified instructors who possess both theoretical knowledge and practical entrepreneurial experience, which limits the effectiveness of teaching [9]. As a result, traditional large-scale, classroom-based models of entrepreneurship education are often ill-equipped to provide the necessary resources, leading to a growing disconnect between academic content and the evolving demands of the entrepreneurial ecosystem [5][10].

Artificial Intelligence (AI)-empowered self-learning systems have emerged as a promising solution to address the limitations of traditional entrepreneurship education. Recent studies emphasize AI's potential to enhance personalized learning, bridge the gap between theory and practice, and provide scalable solutions tailored to diverse student needs [11][12]. AI technologies create a more engaging and dynamic learning environment, allowing students to progress at their own pace while receiving customized guidance [13]. By utilizing AI in entrepreneurship education, students can receive personalized support from ideation to business plan development, with real-

---

time feedback, adaptive learning paths, and tailored guidance—capabilities that traditional models often fail to provide.

This study aims to investigate the design requirements for an AI-driven self-learning system to support business plan development. Specifically, it seeks to identify the essential features and functionalities required to facilitate the development of business plans while addressing the broader pedagogical challenges within entrepreneurship education. Ultimately, this research provides a critical examination of the integration of advanced technologies with entrepreneurship education, aiming to equip students with the necessary mindset and competencies to effectively navigate the evolving entrepreneurial landscape.

## II. LITERATURE REVIEW

### A. AI in Entrepreneurship Education

AI is increasingly recognized for its transformative impact on entrepreneurship education by personalizing learning experiences and connecting theory to real-world applications [7][13]. Intelligent tutoring systems (ITS) and adaptive learning platforms effectively enhance student engagement by providing tailored learning paces, real-time feedback, and personalized pathways, which are crucial for the diverse needs of entrepreneurship students [11][12]. Moreover, AI supports self-regulated learning through continuous, personalized guidance, empowering students to develop critical thinking, decision-making, and problem-solving skills that are vital for entrepreneurial success [14].

In the context of entrepreneurship education, AI applications extend beyond traditional learning tools to offer immersive, hands-on experiences. One key application is the use of AI-driven simulations that allow students to test their business ideas in virtual environments, replicating the uncertainty and risks of real-world entrepreneurship. Such simulations provide experiential learning opportunities, preparing students for the challenges they may face in actual entrepreneurial ventures [5]. In addition to simulations, AI is increasingly leveraged to provide data-driven market insights, enabling students to refine their business strategies and models in real-time. These systems integrate market analytics with business planning, offering immediate feedback to students and enhancing their decision-making capabilities [15]. By enabling real-time adjustments, AI helps students sharpen their business acumen and improve their problem-solving skills [7].

Another significant AI application in entrepreneurship education is the development of virtual business coaches or advisors. These AI systems simulate the feedback loops entrepreneurs would typically receive from mentors or investors, providing personalized advice and encouraging reflective thinking. These systems guide students through the stages of business plan creation, fostering a deeper understanding of the complexities of entrepreneurship and the decision-making processes involved [16]. Additionally, AI has been integrated into gamification platforms, where students participate in entrepreneurship challenges that replicate real-world market conditions. These platforms promote interactive learning and foster an entrepreneurial mindset through competition and hands-on engagement, which enhances students' readiness for entrepreneurial endeavors [17].

Current AI applications in entrepreneurship education, though making progress, remain fragmented and non-integrated, lacking alignment with the classroom context and the processual, incremental, and holistic nature of entrepreneurial learning. Moreover, these applications are often devoid of robust pedagogical frameworks and fail to critically engage with or advance the foundational paradigms of entrepreneurship education. These gaps highlight the need for further exploration into how AI can better empower entrepreneurship education with enhanced pedagogical agency.

### B. Gaps in Business Plan Current Tools

Despite the availability of various business plan creation tools such as LivePlan, Bizplan, and Enloop, these platforms often fail to support the broader objectives of entrepreneurship education. While they help students generate standardized outputs based on predefined templates, these tools primarily emphasize technical aspects—such as financial projections and business structure—at the expense of nurturing the deeper learning processes essential for entrepreneurial success [2].

Moreover, existing tools typically lack personalized feedback and adaptive learning paths, which are crucial for addressing the diverse needs of students in entrepreneurship education [18]. Without real-time feedback and interactive support, students miss the opportunity to refine their ideas and strategies through practical insights and iterative learning [3]. The dynamic and uncertain nature of business ventures requires students to continuously adapt and evolve their strategies, an aspect often overlooked by current tools [19].

Additionally, the pedagogical value embedded in the process of business plan development is often underemphasized. While these tools focus on the technical aspects of creating business plans, they fail to foster creativity, reflective thinking, iteration, and the development of key entrepreneurial competencies such as resilience and adaptability [3]. As a result, students may produce technically sound, one-time business plans that lack the depth of strategic thinking, critical analysis, and authentic skills necessary for success in the real world [5].

Future research should explore how AI-empowered systems can address these gaps by providing a more personalized, interactive, and reflective approach to business plan development. These systems could offer real-time feedback, adaptive learning pathways, and immersive simulations to better prepare students for the complex and dynamic challenges of entrepreneurship. By integrating these features, AI tools could not only assist students in crafting business plans but also scaffold them cultivate the entrepreneurial mindset and competencies that are critical to achieving real-world success.

## III. METHODOLOGY

### A. Research Design

This study adopts a qualitative research approach to explore the system design needs for AI in entrepreneurship education. Qualitative methods are particularly well-suited for this research, as they enable a nuanced understanding of the diverse and context-specific needs of key stakeholders, including educators and students [22]. By capturing detailed insights into their experiences and expectations, this approach provides valuable input for addressing the complex challenges associated with AI integration in educational settings [23].

Semi-structured interviews were chosen for data collection, as they effectively uncover user needs and identify opportunities for improvement by facilitating open-ended dialogue and allowing deeper probing into specific issues [24]. Based on the identified gaps, the interview structure focused on three themes: (1) Mastery of Business Plans, examining participants' understanding and learning needs to pinpoint areas where AI tools can provide targeted support; (2) Core Goals of Entrepreneurship Education, exploring how AI tools can align with the broader educational objectives; and (3) System Design Features, investigating participants' technical expectations for AI tools (see Table 1 in Appendix).

*B. Data Collection*

Data were collected through semi-structured interviews, which offered rich qualitative insights into participants' needs and expectations. The study engaged two key stakeholder groups: (1) Educators, to understand the pedagogical requirements, current practices, and challenges of integrating AI into teaching practices; and (2) Students, to gather feedback on their experiences with AI tools, their learning challenges, and their expectations for personalized support in entrepreneurship education. Table 2 in the Appendix presents a summary of participant demographics and roles, providing an overview of the diverse perspectives.

*C. Data Analysis*

The data collected through semi-structured interviews were analyzed using thematic analysis, a rigorous method for identifying and interpreting patterns within qualitative data [25]. Transcripts were carefully reviewed and initially coded to identify segments relevant to pedagogical challenges, system expectations, and learning needs. These codes were then organized into overarching themes aligned with the study's focus areas: mastery of business plan development, entrepreneurship education goals, and system design features, with sub-themes developed to capture more specific insights. Following a process of review and refinement, the themes and sub-themes were interpreted in relation to the identified gaps and challenges in entrepreneurship education, generating actionable insights to guide the AI system design. *Table 3* presents the coding schemes and results, offering a comprehensive summary of the analysis.

IV. EMPIRICAL FINDINGS: SYSTEM DESIGN NEEDS

By analyzing the perspectives of educators and students, the study identifies critical areas where AI tools can address existing challenges and enhance the learning experience. The findings are structured around three core themes: mastery of business plan development, alignment with entrepreneurship education goals, and technical expectations for system design. These insights form the foundation for proposing a user-centered AI system that supports both pedagogical and practical objectives in entrepreneurship education.

*A. Mastery of Business Plan*

*1) Structure*

The structure of the business plan, as outlined in the guidance book from Cambridge University comprises nine sections: *Summary, Brief History, Product or Service, Markets and Competitors, Marketing Plan, Production and Operations, Management and Objectives, Finance and Risk, Return*, and *Exit* [26]. In contrast, based on interviews with four educators and eight students, the structure for startups business plans in China omits sections such as *Brief, History, Risk, Return*, and *Exit*. Instead, it places a greater emphasis on the analysis of user pain points, which identifies the core problems that the product or service aims to solve and demonstrates a deep understanding of the needs of the target users (E1, E2, E3, S1). As E1 noted, "Problem statement is often much harder than solving the problem itself." For example, university students who want to solve a problem for elderly people may find it difficult to develop truly need-based products or services if they are not even familiar with the daily needs of their grandparents (E3). This illustrates that field research helps people gain a deeper understanding of real needs, allowing them to more accurately define actual problems, and avoid the mismatch between solutions and real needs that might arise from subjective assumptions. E3 points out that:

*Problem Statement* should not be imagined out of thin air; instead, it should be developed from the ground up based on real user research and in-depth communication with actual groups in need.

To ensure logical coherence in each section of the business plan, the following structure is recommended: First is to clearly state the idea background, followed by an explanation of how user needs were identified and the evidence supporting these findings (E1). Next, it should highlight the market demand and specify the market share that is aimed to be captured (S1, S2, S3). Then, outline the implementation plan and articulate how these needs connect with the proposed service, product, or idea. If the idea is well-developed, explain the profit model, detailing the entire process from conception to product launch and market profitability (S1). For potential investors, demonstrate how profitability should be achieved, including short-term and long-term gains, payback period, high-income expectations, and strategies for navigating through business challenges to ensure sustainable development (E3, S1). Each section should have clear key points, with logical coherence between paragraphs and sentences. Avoid imbalances such as being overly detailed at the beginning or overly vague overall (E1).

*2) Language*

The language in a business plan should be clear, smooth, and investor-friendly while incorporating an engaging and appealing tone (E1). Structuring the content as a compelling narrative enhances readability and captures investors' interest, motivating them to delve deeper into the project's details and potential (S1). At the same time, it is crucial to address the challenge of transforming an abstract idea into concrete content. Drawing on E3's insights, the focus should be on illustrating the process of addressing a pain point. This requires accurately identifying the pain point and vividly describing it, making the idea more tangible and reinforcing the plan's logic and persuasiveness.

E3 further emphasized the enduring importance of language in the AIGC era, noting that language is not merely a means of communication but a reflection of underlying logical thinking. "In the AIGC age, the clarity of language remains crucial in entrepreneurial education because it mirrors the clarity of thought. If a person's writing is unclear, it suggests that their ideas are disorganized, and they are unlikely to produce a clear and mature business plan." This insight underscores the critical role of language in conveying coherent and well-structured ideas. In this context, AIGC tools can serve as valuable aids, helping individuals refine and optimize their ideas through enhanced linguistic clarity and logical coherence.

*3) Context*

Developing a compelling business idea requires careful alignment with broader contexts, rigorous validation, and effective communication. These elements ensure that the idea is both relevant and persuasive to potential investors and stakeholders.

Idea formation is adjusted based on extensive research, such as field and market research, guided by the direction anticipated by national expectations. The E1, with extensive experience in business competitions and entrepreneurship, emphasized that verifying whether an idea fits within key areas such as large-scale medical robotics, low-altitude economy, high-end manufacturing, and cultural tourism is crucial. He emphasized that it's crucial to check if the project fits the key areas currently supported by the government. For example, in competitions like the "Innovation and entrepreneurship" contest or "Internet Plus," the initial assessment involves confirming whether the idea matches these national priorities. This alignment not only enhances the idea's relevance but also increases its chances of receiving support and funding.

A solid business idea must also be grounded in thorough market validation. This includes conducting market analysis to assess profitability and market size, ensuring that the product addresses an unmet need. Almost all interviewees mentioned this as a critical component. It also evaluates whether the team has the necessary resources and expertise to meet the demand (S1, S2). Financial projections, including costs and potential profits, are essential for demonstrating the business's viability (S1). These steps, grounded in research, clarify the business idea's value and its capacity to generate growth and attract investment.

Even the same idea will have different contexts due to different listeners. When presented to different groups or investors, the idea should be tailored to align with their specific needs and interests to make a greater impact. E3 emphasizes that entrepreneurship is inherently uncertain, with no guaranteed formula for success. Even if one follows a proven structure or format from previous successful business plans, it doesn't ensure that investors will fund the project. S1 also used his experience in business competitions to demonstrate that different audiences require different presentation styles. In many cases, it may be necessary to include an extra slide to capture the audience's attention or to summarize with a single, impactful slogan. Most educators highlight the importance of diversity in perspectives, as different stakeholders may have varying expectations and concerns (E1 and E3).

*B. Entrepreneurship Education Needs*

*1) Understanding*

The interviews revealed several key insights regarding the role of business BP writing in student entrepreneurship education. BP writing helps students transform their creative, abstract ideas into concrete, actionable plans, clarifying their entrepreneurial vision and enhancing their ability to articulate and justify their ideas' potential profitability and scalability (E1, S1).

The process also prompts students to critically evaluate their ideas against market demands and assess the viability of their ventures. This step-by-step approach allows them to address critical questions related to market demand, profitability, and growth potential, thereby preparing them for real-world entrepreneurial challenges (E1, E3, S1).

Moreover, BP writing serves as a scaffold for developing a business mindset. It guides students to think critically about their ideas, structure them clearly, and define how they will meet market needs. This process fosters a focus on growth, profitability, and scalability—essential elements for entrepreneurial success (E1, E3, S1).

Additionally, by following a structured BP format, students learn to break down their ideas into specific, actionable steps and identify the necessary resources and strategies for implementation. This skill is crucial for navigating the entrepreneurial journey and building a foundation for sustainable business growth (E2).

These findings highlight the importance of BP writing as an educational tool that enhances students' understanding of entrepreneurship and equips them with practical skills and a mindset conducive to entrepreneurial success.

*2) Guidance*

Provides specific writing steps and examples for each module in the structure, as well as guidance on writing effective prompts with the assistance of LLM. For example, for the Financial Projections section, students can use prompts like *"Help me outline the expected revenue streams and cost structure for a new educational tech startup."*(S7). There is some ambiguity regarding whether LLMs should provide industry-specific data or insights. While educators (E1, E3) believe that LLMs should focus on guiding students in structuring and articulating their business plans rather than providing industry-specific data, students often expect LLMs to offer relevant information to ease their writing process (S4, S6, S7). Educators emphasize the importance of developing research skills by independently sourcing data, as this mirrors real-world entrepreneurial challenges (E1, E3).

In terms of analysis tools, there are several recommendations for writing each module. For risk assessment, tools like TAM (Total Addressable Market) and TPM (Total Product Market) are valuable in evaluating market size and product positioning (S3). Visual aids such as bar charts and line graphs can help students present market trends and potential growth (S2,). While SWOT analysis remains useful for assessing internal and external factors, integrating it with TAM and TPM can provide a more comprehensive perspective (S1, S2, S4). For market research, students can leverage survey tools like Google Forms or SurveyMonkey, with qualitative techniques such as interviews, guided by open-ended questions and the "Five Whys" technique to uncover deeper insights (E2). Finally, for interview-based research, transcription tools like mind mapping tools can help students organize and synthesize insights, facilitating a structured approach to understanding customer needs and refining business strategies.

These insights have significant implications for designing AI system tools. They highlight the need for AI tools to strike a balance between providing structured guidance and encouraging independent research. By integrating these capabilities, AI systems can better support students in developing essential entrepreneurial competencies.

*3) Contextualization*

Using LLMs to simulate roadshow scenarios for BPs was identified as a valuable training method for students. Educators emphasized the importance of students collecting data on investor preferences and setting up different scenarios based on this information. One educator (E2) highlighted that analyzing successful investment presentations helps students identify which aspects particularly appeal to technology investors. "This process not only helps them understand investor preferences but also develops their ability to actively gather and apply information in real business contexts," E2 explained.

When discussing how students adapt their presentations to different investor profiles, several students emphasized the importance of flexibility. For example, one student (S4) mentioned the need to tailor their pitch to match investor priorities: "When presenting to investors focused on social impact, we learned to emphasize the product's contribution to specific communities, like the elderly. It's crucial to align our message with their values." Another student (S7) added that these simulations taught them to think critically about tailoring their BP content. "Leveraging information gaps as a strategic resource in entrepreneurship is key," S7 noted.

Simulating the long-term effects of entrepreneurship was also seen as crucial for understanding its complexity. One educator (E3) proposed that students gather initial user data, run small-scale pilots, and then expand the user base to observe the impact. "This approach allows them to experience various entrepreneurial decisions and challenges in a virtual sandbox environment," E3 explained. Modern virtual sandboxes allow students to see the long-term impact of their decisions in different scenarios by introducing roles such as virtual investors, users, and markets. By setting different time spans (such as 5 or 10 years), students can see how the financial forecasts and impact assessments evolve. As one student (S6) observed, "Seeing the long-term impact of our decisions helps us understand whether our strategies will succeed or fail. It's like a safe space to experiment and learn from our mistakes."

These findings suggest that AI tools should prioritize features that support real-world scenario simulations and long-term impact analysis through interactive dialogues. By integrating these capabilities into an interactive system, students can engage in dynamic conversations to experiment, adapt, and refine their strategies. This approach would not only enhance their entrepreneurial preparedness but also provide a more personalized and responsive learning experience.

*4) Engagement*

An important aspect of improving student participation is ensuring that the role of LLMs is not to foster over-reliance or complacency, but to enhance students' understanding of the educational process and stimulate active thinking and learning, as emphasized by E1. By inputting data into LLM in educational practice, students can initially form business ideas and dynamically adjust and optimize these ideas through the process of constantly supplementing data. This method not only enables students to obtain real-time feedback and guidance but also deepens their understanding of the process of creative formation and adjustment process, according to E2.

Beyond merely providing feedback, LLMs can significantly enhance the vividness of teaching by simulating real-world situations, making it easier for students to integrate into the learning environment and stimulating their interest in participation, as noted by E2 and E3. These simulation scenarios allow students to experiment with different decisions in a safe environment, better preparing them for real-world challenges.

At a deeper level, this interactive and immersive learning experience helps students understand the core value of education by connecting theoretical knowledge with practical application. As E3 highlighted, "When students see the immediate impact of their decisions in simulated scenarios, they become more engaged and motivated to learn." This connection between theory and practice not only improves their enthusiasm but also fosters a sense of ownership and relevance in their learning journey, ultimately enhancing their overall engagement.

*5) Interaction*

Multiple rounds of communication between users and LLMs are crucial for refining and enhancing creative concepts. Many students highlighted the importance of iterative discussions in shaping their ideas. For example, S1 mentioned, "Discussing my business idea with friends helped me see its strengths and weaknesses from different angles." Similarly, S3 and S7 noted that these exchanges provided valuable insights into market demand and the feasibility of their innovations. As S6 put it, "Talking through my ideas with others made me realize what aspects needed more work and which ones were truly unique."

Simulating interactions with investors was another key strategy mentioned by educators (E1, E3) as an effective way to improve students' ideas. These mock conversations allow students to respond to potential investors' questions and feedback, helping them adjust their plans to meet investor expectations. E1 emphasized, "When students practice presenting their ideas to simulated investors, they learn to anticipate tough questions and refine their pitches accordingly." This process not only strengthens their entrepreneurial thinking but also enhances their practical skills by identifying potential gaps.

These findings suggest that an AI system should be designed to facilitate interactive and iterative exchanges, mimicking both peer discussions and investor interactions. By incorporating features that support multi-turn dialogues and simulated feedback, the system can help students test and refine their ideas in a dynamic and supportive environment.

*6) Reflection*

Reflective practice is often overlooked in system design, yet it is a fundamental and deeply transformative element in entrepreneurship education. One educator (E1) emphasized that reflection is key to measuring whether students have truly learned knowledge and abilities. "It's not just about proposing ideas; it's about understanding how those ideas evolved and what skills students developed along the way," E1 noted. Integrating reflection sessions into each module helps students deeply examine their learning process and progress (E1, S4, S5, S6).

One type of reflection is skills reflection, which focuses on the entrepreneurial skills students develop during the learning process (E2). To help students track their growth, a checklist of key entrepreneurial skills can be provided at the start of the course, allowing students to self-assess based on these skills. Throughout the course, students can periodically evaluate themselves against this checklist, reflecting on their progress. At the end of the course, students can use a learning roadmap to clearly see how their abilities have changed, identifying which skills have significantly improved and which areas still have gaps. This type of reflection helps students focus on the development of specific skills and assess their progress through real-life growth.

Another type of reflection is critical thinking reflection, which focuses on the key incidents students experience in real-life situations. Some educators (E3, E4) and students (S4, S5, S7) suggested that LLMs could provide feedback on students' approaches, allowing them to reflect on how changes improve outcomes. "When students encounter unexpected challenges or successes, those moments become powerful opportunities for reflection," E3 noted. By examining these experiences, students can identify which skills helped them navigate challenges and which areas hindered their performance (E1, E2).

Incorporating these reflective practices into AI-driven systems not only enhances students' critical thinking and self-awareness but also adds a layer of educational depth and empathy to the learning experience. By prompting students to reflect on both their skills and emotional responses, AI tools can create a more holistic and impactful learning environment—one that is not only effective but also emotionally resonant and supportive.

C. **System Design Features**

*1) Customized Content*

For effective learning, the system should provide customized content based on students' unique ideas and the diverse roles they play during the learning process—such as entrepreneurs, investors, or customers. Interviewees emphasized that the system needs to offer tailored guidance and feedback to support each student's specific needs.

E1 suggested that the system should dynamically generate content based on students' input and feedback to help them better understand and refine their business plans. S1 highlighted the importance of personalized suggestions, noting, "Customized content is very important. The system should provide personalized suggestions based on my needs and progress, helping me to view my plan from different angles, especially in the context of my professional background."

These insights indicate that the system should be adaptive and responsive, offering personalized support that enhances students' understanding and development of their BPs.

*2) Adaptability*

Building on the need for customized content, the system must also be adaptable to the diverse learning stages and evolving needs of students. E1 emphasized that adaptability is crucial, allowing the system to adjust to different learning environments and provide a truly personalized experience. E4 further noted that the system should offer step-by-step, targeted guidance, progressing from the initial stages of BP writing to more in-depth business logic analysis. This approach ensures that students receive the right support at the right time, based on their individual progress and changing needs.

Students themselves highlighted the importance of this adaptability. S2, S5, and S7 pointed out that the system should address practical challenges at each stage of the BP development process, helping them navigate complexities and solve real-world problems. By being responsive to students' evolving needs, the system can go beyond merely providing personalized content—it can actively facilitate continuous growth and learning, making it a dynamic and essential tool throughout their entrepreneurial journey.

*3) Real-time Interaction*

All educators believe that the system should provide real-time interactive capabilities to support students during the BP writing process. E4 emphasized the importance of timely feedback, noting that it allows students to refine their ideas and make adjustments on the fly. "When students receive immediate feedback, they can quickly identify weaknesses and improve their plans," E4 explained.

S1 also highlighted the value of real-time interaction, saying, "I find real-time interaction very useful, especially when I encounter a bottleneck. The system can provide instant feedback and adjustment suggestions to help me move forward." This immediate support is crucial for helping students navigate challenges and stay engaged in the learning process. As highlighted by E1, S4, and S5, the system should offer real-time feedback to help students continuously adjust and optimize the content of their plans.

*4) Usability*

Students consistently emphasized the importance of a simple and clear interface, as well as an intuitive operation process. A streamlined design allows them to focus more on content creation without being bogged down by complex functionalities. As noted by S2, S3, S4, S6, and S7, a user-friendly interface saves time and enhances the efficiency of learning and writing.

E3 also highlighted that the system should have an intuitive interface and smooth operation process to ensure efficient use. "An easy-to-use system means students can quickly get started and spend more time on developing their ideas rather than figuring out how to use the tool," E3 explained. This emphasis on usability underscores the need for a design that prioritizes accessibility and efficiency, ultimately supporting a seamless learning experience.

*5) Aesthetics.*

The visual design of the interface plays a crucial role in enhancing students' learning experience and their willingness to engage with the system. E4 emphasized the importance of good visual effects, noting that an attractive interface can significantly boost students' interest and provide a more enjoyable learning experience.

S1 echoed this sentiment, saying, "If the system interface is designed to be both beautiful and practical, I will be more willing to spend time on it and increase my interest in learning." Other students (S3, S5, S7) also highlighted that a visually appealing interface not only makes them more likely to use the system but also helps them stay focused and motivated during use.

Overall, an aesthetically pleasing design is more than just a superficial feature; it can actively contribute to a positive learning environment and sustained user engagement.

## V. System Design Exploration

### A. Pathways for Implementation

Figure 1 illustrates the interface between users and the LLM, highlighting three key goals: idea discussion, business plan (BP) writing, and pitching simulation. The design is driven by the need to address common challenges identified through interviews, such as the need for adaptability, real-time interaction, and user-friendly navigation. By incorporating structured templates, real-time feedback, and personalized writing suggestions, the interface aims to support students in refining their ideas and developing their BPs effectively. Additionally, interactive elements like simulated investor feedback and progress tracking are included to enhance engagement and ensure the system remains intuitive and visually appealing.

In Figure 1, section 1 is structured to guide students through three key functions: idea formation, business plan writing, and pitching simulation. Section 2 provides step-by-step assistance in writing the business plan, with a clear structure that allows students to focus on one section at a time. Each section includes targeted guidance, example content, and tips to help students. Section 3 provides ongoing guidance throughout the process, with the system offering real-time suggestions, tips, and personalized feedback based on students' inputs. Section 4 includes an interactive dialogue feature, where students can input prompts and engage in conversations with the system

### B. Technical Challenges

The proposed system faces several technical challenges that stem from its ambitious goals and the complexity of its requirements. These challenges highlight the gap between the aspirational features of the system and the current limitations of existing technologies.

One of the primary challenges is the implementation of multimodal LLM (Large Language Models), which requires integrating different types of data, such as text, images, and possibly voice. This integration demands more advanced technical capabilities, as the system needs to process and interpret complex multimodal inputs in real-time. Ensuring that the AI can handle nuanced, context-dependent user requests while maintaining accuracy and relevance is a significant hurdle.

Another challenge is the real-time feedback system, which is essential for guiding students through the process of business plan writing and idea formation. The time required for generating real-time feedback is heavily influenced by the size of the AI model being used. Larger models may offer more accurate and context-aware responses but come with increased processing times. This creates a tradeoff between response speed and model performance, and it is difficult to improve response times solely through system design without compromising the quality of the feedback.

The user interface (UI) design also presents significant challenges. The system must provide an intuitive and accessible experience for a diverse user base, which may include individuals with varying levels of technical proficiency. Balancing ease of use with the functional complexity required for business plan writing, idea generation, and real-time interactions is not straightforward. Moreover, the interface must cater to different learning styles and provide a seamless experience across various devices and platforms. Creating such a flexible yet user-friendly UI is a complex task that requires careful design consideration.

In conclusion, while the proposed system aims to offer personalized, real-time support for students in the process of entrepreneurship learning, it faces substantial technical hurdles. These challenges include the integration of multimodal LLM, balancing real-time feedback with model size, and designing an intuitive UI that caters to a broad user base. Overcoming these issues requires both technical innovation and a deep understanding of user needs to create a system that is both functional and accessible.

## VI. Conclusion and Implications

This study deepens our understanding of how AI tools can be integrated into entrepreneurship education to address both long-standing challenges and emerging opportunities. Prior research has emphasized the importance of structured guidance in entrepreneurial learning, particularly in areas such as business plan development and fostering entrepreneurial mindsets [2] [18]. Building on these foundations, the findings presented here not only confirm these priorities but also identify nuanced, practical needs for AI-driven solutions. These insights pave the way for a paradigm shift in entrepreneurship education, where technology serves as a dynamic scaffold to enhance learning outcomes, creativity, and strategic thinking while supporting foundational human elements of the entrepreneurial process.

### A. Discussion and Insights of AI-Scaffolding Business Plan Development and Entrepreneurship Education

One key insight is the necessity of scaffolding students' mastery of business plan development. Consistent with [3], the findings underscore that students often struggle with structuring and articulating their ideas, particularly in transforming abstract creativity into actionable frameworks. The system must go beyond supporting writing, language, and content to also assist with foundational entrepreneurial thinking and idea formation.

This aligns with Kourilsky & Walstad's [19] emphasis on fostering foundational entrepreneurial thinking as a precursor to successful business planning, as well as Nabi et al.'s [18] recognition of the importance of aligning educational support with entrepreneurial challenges. Scaffolding should help students understand market dynamics, identify trends, and bridge gaps in analytical skills, ensuring that business plans are both creative and practically aligned with real-world demands.

Moreover, the importance of guiding the business idea formation by AI should be emphasized. An AI-driven system must address a common issue among technical entrepreneurs—prioritizing technology development over identifying market needs. This misaligned approach, often detached from mission-driven motivation and user-centric perspectives, leads to ventures that fail to meet real-world demands. Echoing Brem [5], the findings suggest that AI systems should emphasize early-stage idea development by grounding entrepreneurs in understanding user needs and market dynamics, ultimately reshaping entrepreneurial practices. By promoting idea formation driven by purpose and societal impact, such systems can guide entrepreneurs to align their ventures with market realities while inspiring foundational innovations that contribute to global priorities. This dual focus on market alignment and societal impact positions AI systems as catalysts for sustainable and impactful entrepreneurial behaviors [20][21].

Entrepreneurship education, with AI-driven systems, can address persistent challenges by offering more personalized and transformative learning experiences. By focusing on understanding, guidance, contextualization, engagement, interaction, and reflection, these systems have the potential to significantly enhance entrepreneurial learning. A key issue in entrepreneurship education is students' struggle to connect the technical components of business plans with their strategic purpose. AI systems can help bridge this gap by explaining both the function and rationale behind each section, fostering critical thinking about the strategic coherence of business plans.

Traditional entrepreneurship education often relies on generic guidance that fails to adapt to learners' specific needs. As Morris et al. [7] note, effective entrepreneurial learning requires adaptive scaffolding that addresses unique challenges faced by students at different stages of the business planning process. AI systems provide adaptive, step-by-step instructions tailored to individual progress, enhancing both technical quality and confidence. Contextualization is crucial for bridging theoretical knowledge and practical application. AI-driven tools can simulate real-world scenarios and integrate market data, improving students' ability to refine their plans and address market needs. Engagement, often a challenge in traditional methods, is improved by AI systems that require active participation, making students co-creators of their learning process. Finally, AI systems support reflection by prompting students to evaluate assumptions, critique strategies, and iterate on their plans, fostering critical thinking and resilience—key qualities for entrepreneurial success.

By addressing these dimensions, AI-driven systems provide solutions to the fundamental limitations of traditional entrepreneurship education. They enable deeper conceptual understanding, offer tailored and adaptive guidance, situate learning within practical contexts, foster active participation, and promote critical reflection. These capabilities contribute to a new paradigm for entrepreneurship education—one that empowers learners to engage dynamically with entrepreneurial processes, bridging the gap between theoretical knowledge and practical application while fostering the adaptability, creativity, and resilience necessary for real-world success.

*B. Future Development and Research Directions*

This study highlights several technical challenges in refining AI-driven systems for entrepreneurship education. A key challenge is developing systems that offer personalized content and learning paths tailored to diverse student needs. This requires advances in AI-driven adaptive learning algorithms [16]. Real-time interaction remains a hurdle, as AI systems must deliver timely, context-specific feedback without losing depth, necessitating improvements in natural language processing [17]. Additionally, usability and aesthetic design are crucial to ensuring that AI systems are intuitive and engaging [15].

While AI can enhance entrepreneurship education by scaffolding self-regulated learning, it cannot replace essential human elements. Mentorship, hands-on practice, and collaboration are vital for developing practical entrepreneurial skills [18]. AI should complement these human-centered elements, enhancing the learning experience. Hybrid learning models that integrate AI for foundational knowledge and human-led courses for experiential learning could balance self-directed study with collaborative practice [5]. Embedding AI in broader educational ecosystems can also support lifelong entrepreneurial learning, helping learners adapt to evolving market dynamics. Furthermore, embedding AI into broader educational ecosystems can enable lifelong entrepreneurial learning, helping learners continuously adapt to evolving market dynamics and societal needs.

In conclusion, while AI offers remarkable potential to revolutionize entrepreneurship education, its true value lies in enhancing and complementing human-led practices rather than replacing them. By fostering collaboration between AI systems and traditional instruction, the future of entrepreneurship education can evolve into a holistic model that combines personalized, technology-driven learning with experiential, practice-oriented approaches. This integrated perspective highlights the dual role of AI systems: as tools that address specific educational deficiencies and as catalysts for fostering deeper, more adaptable entrepreneurial behaviors in a rapidly evolving global context.

VII. APPENDIX

A.  Table 1: Interview Structure and Focus

| Theme | Focus |
| --- | --- |
| Mastery of Business Plans | Understanding participants' knowledge gaps and learning needs regarding business plan creation, and identifying how AI can provide step-by-step guidance, such as structuring ideas or evaluating feasibility. |
| Core Goals of Entrepreneurship Education | Exploring the role of AI in supporting key educational outcomes like critical thinking, problem-solving, and practical application of entrepreneurial concepts, ensuring the tools complement the overall learning objectives. |
| System Design Features | Investigating specific technical expectations, such as user interface preferences, feedback mechanisms, real-time assistance, and adaptability to different stages of business plan development. |

*B. Table 2: Interviewers' Demographics*

| Number | Demographics | Backgrounds |
|---|---|---|
| E1 | Educator | Lecture on innovation and entrepreneurship education |
| E2 | Educator | Manager of the innovation and entrepreneurship project incubation |
| E3 | Educator | Lecture teaching innovation and entrepreneurship education |
| E4 | Educator | Lecture with professional human-centered interaction |
| S1 | Student | Entrepreneurial experience and professional business |
| S2 | Student | Data Analysis, Business. Have a startup |
| S3 | Student | Data Analysis, Have a startup |
| S4 | Student | entrepreneurial experience; Policy Education |
| S5 | Student | entrepreneurial experience; skilled programmer |
| S6 | Student | entrepreneurial experience; Chip Manufacturing |
| S7 | Student | entrepreneurial experience law |
| S8 | Student | entrepreneurial experience; skilled programmer |

*C. Table 3: Coding Themes, Sub-themes, and Key Phrases*

| **Themes: Mastery of Business Plan Writing** | | |
|---|---|---|
| **Sub-themes** | **Definition** | **Key Phrases** |
| Structure | The organization of a business plan to ensure logical coherence | pain point, problem statement, logically coherent |
| Language | use storytelling to create an engaging narrative | clear, engaging, and investor-friendly, effectively |
| Content | Idea formation is refined through discussion and research, with varying presentations and emphasis | Extensive research, idea refinement, Tailored presentation |
| **Themes: Core Goals of Entrepreneurship Education** | | |
| **Sub-themes** | **Definition** | **Key Phrases** |
| Understanding | helps students grasp the key aspects of entrepreneurship | Entrepreneurship key points, real-world challenges |
| Guidance | Provide clear steps, examples, and tools for each module | writing steps, structured approach, effective prompts |
| Engagement | Encourage active student participation by using LLM | Active involvement, student-centered approach, real-time feedback |
| Contextualization | using LLM to simulate real-world business scenarios | Simulate roadshow scenarios, Investor preferences, long-term impact simulation |
| Interaction | iterative communication between students and LLM | Multiple rounds of communication, refine creative concepts, Discussions with friends |
| Reflection | the process of evaluating and thinking deeply about the knowledge and skills students have acquired | Self-evaluation, learning from mistakes, Self-assessment, Internal reflection |
| **Themes: System Design Features** | | |
| **Sub-themes** | **Definition** | **Key Phrases** |
| Customized Content | Tailor the system's output and feedback based on individual students' needs | Personalized learning paths, individualized feedback, targeted support, tailored suggestions" |
| Adaptability | The system's ability to adjust to students' varying needs | Flexible learning paths, adjusting to student progress, Tailored step-by-step progression |
| Real-time Interaction | Enable immediate feedback and adjustments | Instant feedback, interactive guidance, real-time collaboration" |
| Usability | the system has an intuitive, easy-to-navigate interface | User-friendly interface, easy navigation, intuitive design, quick setup" |
| Aesthetics | The visual appeal of the system's interface | Visually appealing design, engaging layout, pleasant user experience |

D. Figure 1: UI Design.

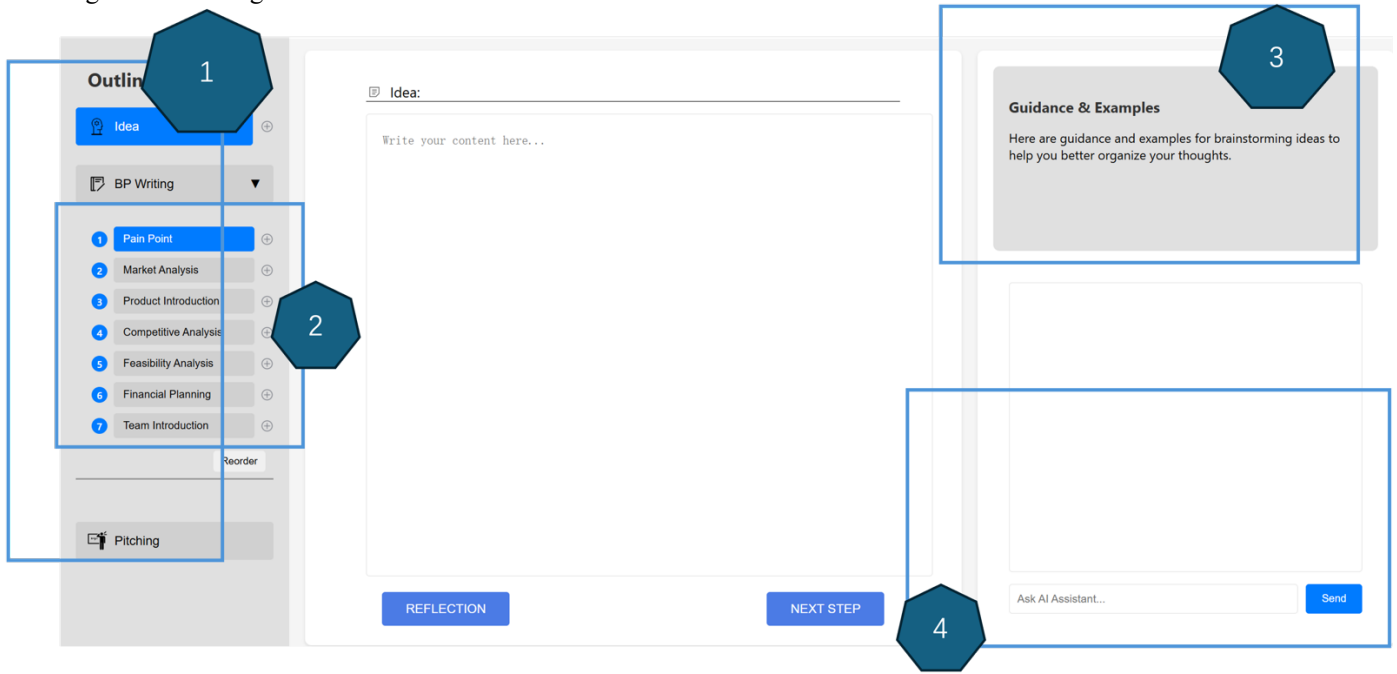